# Domain Matching Epitaxy of Ferroelectric $Hf_{0.5}Zr_{0.5}O_2(111)$ on $La_{2/3}Sr_{1/3}MnO_3(001)$

Saul Estandía,[1] Nico Dix,[1] Matthew F. Chisholm,[2] Ignasi Fina,[1] Florencio Sánchez[1, *]

[1] Institut de Ciència de Materials de Barcelona (ICMAB-CSIC), Campus UAB, Bellaterra 08193, Barcelona, Spain

[2] Center for Nanophase Materials Sciences, Oak Ridge National Laboratory, Tennessee 37831-6064, USA

Epitaxial ferroelectric $HfO_2$ films are the most suitable to investigate intrinsic properties of the material and for prototyping emerging devices. Ferroelectric $Hf_{0.5}Zr_{0.5}O_2(111)$ films have been epitaxially stabilized on $La_{2/3}Sr_{1/3}MnO_3(001)$ electrodes. This epitaxy, considering the symmetry dissimilarity and the huge lattice mismatch, is not compatible with conventional mechanisms of epitaxy. To gain insight into the epitaxy mechanism, scanning transmission electron microscopy characterization of the interface was performed, revealing arrays of dislocations with short periodicities. These observed periodicities agree with the expected for domain matching epitaxy, indicating that this unconventional mechanism could be the prevailing factor in the stabilization of ferroelectric $Hf_{0.5}Zr_{0.5}O_2$ with (111) orientation in the epitaxial $Hf_{0.5}Zr_{0.5}O_2(111)/La_{2/3}Sr_{1/3}MnO_3(001)$ heterostructure.

* E-mail: fsanchez@icmab.es

## Introduction

The robust ferroelectricity in doped hafnia and zirconia nanometric thin films, first reported in 2011,[1] is causing huge scientific and technological interest.[2–6] The observation of ferroelectricity was unexpected, since the crystal structure of bulk $HfO_2$ and $ZrO_2$ oxides at room temperature and ambient pressure is non-polar monoclinic (space group number 14, $P2_1/c$).[7–10] Varying temperature and pressure they can be tetragonal (137, $P4_2/nmc$), cubic (225, $Fm3m$) or orthorhombic (oI phase: 61, Pbca or oII phase: 62, Pnma). All these phases are non-polar. However, other metastable polar polymorphs whose energy is close to the stable phases exist, and among them, there are two orthorhombic structures (oIII: 29, $Pca2_1$, and oIV: 31, $Pmn2_1$)[8,9,11], and recently a polar rhombohedral phase (160, R3m) has been also reported.[12,13]

The interest in ferroelectric $HfO_2$ emerges from the full compatibility with CMOS technology of the chemical composition and the deposition processes to obtain polycrystalline ferroelectric $HfO_2$ films.[14] However, the current knowledge on ferroelectric hafnia needs to be improved. In this regard, epitaxial films [15–19] can be useful to reveal the underlying mechanisms governing $HfO_2$ properties, as well as to prototype nanometric $HfO_2$-based devices. Examples of the usefulness of epitaxial films include observation of coercive electric field ($E_c$) scaling with thickness (t) according to the $E_c \sim t^{-2/3}$ dependence (elusive for polycrystalline $HfO_2$),[20] absence of wake-up effect,[21–23] and disentanglement and control of ionic and electronic transport contributions in tunnel devices.[24] Since these significant advances may result crucial for the further improvement of polycrystalline ferroelectric $HfO_2$, the development of epitaxial $HfO_2$ is of major relevance.

Polycrystalline ferroelectric $HfO_2$ are typically obtained by annealing amorphous nanometric films. The mechanisms of formation of the metastable phase are under discussion. Surface and interface energy contributions, and changes in energy caused by lattice strain in the crystallized films, can be a very relevant fraction of the total energy due to the nanometric sizes.[9,25] Moreover, even if the monoclinic phase is thermodynamically stable at room temperature, the transformation from other phase, when the film is cooled, could be kinetically suppressed.[26] A different method to form the metastable phase is epitaxial growth, generally by pulsed laser deposition (PLD) at high temperature. In this deposition process, contrary to the annealing of amorphous films, crystallization of the atoms that arrive from the phase gas occurs in a very short time scale.[27] In the early growth stage, the contribution of the substrate/film interface energy is larger than the bulk energy. The energy of an incoherent interface is much higher than that of a coherent or semicoherent one, and thus a particular polymorph will form if the crystal surface of the used substrate permits (semi)coherent growth.[28] The phenomenon is usually named epitaxial stabilization when the formed phase is not the one having the lower energy in bulk.[29]

Epitaxial ferroelectric doped-HfO$_2$ capacitors were firstly obtained with Y:HfO$_2$ and indium tin oxide electrodes on single crystalline yttria-stabilized zirconia (YSZ) substrates.[15] Later, high polarization epitaxial Hf$_{0.5}$Zr$_{0.5}$O$_2$ (HZO) films, (111)-oriented, were deposited on SrTiO$_3$(001) substrates buffered with La$_{2/3}$Sr$_{1/3}$MnO$_3$ (LSMO) electrodes.[12,21] The ferroelectric phase was proposed to correspond to orthorhombic Pca2$_1$ [21] or rhombohedral R3m [12] phases. The monoclinic phase coexists in these films, and the deposition parameters allow controlling the relative amount of phases and polarization.[20] The ferroelectric capacitors show high polarization, endurance and retention.[21] Epitaxial HZO films were also epitaxially integrated with Si(001), using LSMO as electrode and either YSZ or SrTiO$_3$ as buffer layers.[22,23] On the other hand, the selection of the substrate allows selection of the growing HfO$_2$ phase (pure monoclinic, pure orthorhombic, or a mixture of both), as has been recently demonstrated in HZO/LSMO heterostructures grown on a set of oxide single crystals that exerts varying epitaxial stress.[19]

Epitaxial HZO/LSMO is thus a very useful platform to investigate properties of ferroelectric HfO$_2$ and to fabricate emerging devices such as ferroelectric tunnel junctions. However, the epitaxy mechanism of this functional heterostructure, HZO(111)/LSMO(001), is intriguing and still unknown. Indeed, epitaxy was unexpected considering the symmetry dissimilarity and the large lattice mismatch between orthorhombic HZO(111) on LSMO(001). How does (111)-oriented orthorhombic HZO grow epitaxially on LSMO(001)?

Covalent epitaxy requires crystal matching at the interface between film and substrate (or lower layer). Cube-on-cube epitaxy is common when lattice mismatch is low. Other simple cases, as 45° in-plane unit-cell rotation, are frequent.[30] However, these simple mechanisms of epitaxy are not feasible when substrate and film are more dissimilar. In these cases, films are generally polycrystalline, although some less usual mechanisms can permit epitaxy. Examples include tilted epitaxy [31] and coincidence lattice epitaxy,[32] also called domain matching epitaxy (DME).[33]

We have investigated the mechanism of epitaxy studying HZO/LSMO bilayers deposited on GdScO$_3$(001) (for the sake of simplicity, the scandate substrate is indexed here using pseudocubic setting). GdScO$_3$(001) substrates are the optimal to stabilize the orthorhombic phase, and HZO films on LSMO(001)/GdScO$_3$(001) are free of paraelectric monoclinic phase.[19] Semicoherent interfaces between HZO and LSMO are observed by scanning transmission electron microscopy (STEM). It is found that the large lattice mismatch, around -10% along [-211]HZO/[110]LSMO and around 58% along [0-22]HZO/[110]LSMO, is accommodated by DME. In DME, m lattice planes of the upper layer match with n lattice planes of the lower layer, instead of the one-to-one matching of conventional epitaxy. The formation of domains, with m/n values around 9/10 along [-211] and 3/2 along [0-22] HZO directions, permits the relief of the epitaxial stress and the growth of high crystal quality orthorhombic films. The DME mechanism is effective in stabilizing the metastable orthorhombic instead of the monoclinic phase, and results in high quality epitaxial films, unlike the usually defective films grown by conventional epitaxy when lattice mismatch is moderately high.

## Results and discussion

Figure 1a shows the X-ray diffraction (XRD) θ-2θ scan of the HZO/LSMO/GdScO$_3$(001) sample. The two peaks at low angle correspond to the (001) reflections of the GdScO$_3$ substrate and the LSMO electrode. The peak at 2θ around 30.5° is at the position of the o-HZO(111) reflection. A scan around this peak with longer acquisition time, which is shown in Figure 1a and vertically shifted for the sake of clarity, reveals Laue interference fringes. The fitting of simulated Laue fringes (Supporting Information S1) indicates a HZO thickness of 10 nm. Pole figures were measured to determine the epitaxial relationships. Figure 1b shows the pole figures around GdScO$_3$(111) and o-HZO(-111) asymmetric reflections. There are four GdScO$_3$(111) poles and twelve o-HZO(-111) poles. One single o-HZO variant would produce three (-111) poles; therefore, the presence of twelve poles confirms the presence of four o-HZO epitaxial crystal variants. These epitaxial variants imply that o-HZO grains can grow with either [-211] or [0-22] directions aligned with [110] of LSMO. The four crystal domains are a natural consequence of the three-fold symmetry of (111)-oriented HZO film grown on the four-fold symmetry LSMO(001). Crystal domains form in heteroepitaxy when the symmetry of the film is lower than that of the substrate,[34,35] and they can contribute to relieve elastic energy of the film.[36] In summary, XRD indicates epitaxial growth of orthorhombic HZO, with (111) orientation and absence of monoclinic phase. In agreement with the stabilization of the orthorhombic phase, the film is ferroelectric. The polarization loop (Figure 1c) shows a remnant polarization of around 20 $\mu C/cm^2$ and a coercive field of around 3 MV/cm.

The epitaxial relationship between the crystal variants of o-HZO(111) film and the $GdScO_3(001)$ (or the coherently strained LSMO(001) electrode) are sketched in Figure 2a. It is not obvious how lattice matching between the approximately four-fold and three-fold symmetries of LSMO(001) and o-HZO(111) surfaces occurs. Figure 2b depicts the simulated structures of LSMO and o-HZO across the interface,[37] projected on the HZO plane containing $[111]_{op}/[-211]_{ip}$ (op and ip subindices indicate out of plane and in-plane, respectively) and, correspondingly, the $[001]_{op}/[110]_{ip}$ plane of LSMO. The same structure projected on the in-plane orthogonal direction of HZO, $[111]_{op}/[0-22]_{ip}$, and $[001]_{op}/[110]_{ip}$ of LSMO, is presented in Figure 2c (additional details can be found in Supporting Information S2). The atomic arrangement of LSMO implies an in-plane distance between cation columns parallel to the interface of approximately 2.807 Å, corresponding to the (110) interplanar distance. This distance is determined by the $GdScO_3$ substrate, since the LSMO layer grows fully strained on it.[19] On the other hand, the atomic arrangement of o-HZO, somewhat more complex, has been simulated using VESTA software considering calculated lattice constant for orthorhombic ($Pca2_1$) $HfO_2$.[37,38] Our STEM results are in agreement with the simulations done considering the $Pca2_1$ space group. The HZO cation columns appear positioned along the HZO[-211] direction at a distance of approximately 3.118 Å. In order to analyze the lattice matching of HZO on LSMO, HZO(-111) and LSMO(110) planes were used, see Figure 2b. The lattice mismatch (f) with LSMO (f (%) = 100 x $(d_{LSMO} - d_{HZO})/d_{HZO}$, where $d_{LSMO}$ and $d_{HZO}$ are the horizontal distances between consecutive horizontally aligned cation columns in each material) is -9.97%. The matching of HZO(0-22) and LSMO(110) planes along the orthogonal direction is shown in Figure 2c. The HZO cation columns are at a distance of around 1.776 Å, and the corresponding lattice mismatch with LSMO is 58.0%. Conventional epitaxy (coherent growth, followed by plastic relaxation above a critical thickness) is usual in systems where mismatch is low or moderate, but it is not possible in largely mismatched systems as occurs along HZO[-211]/LSMO[1-10] and along HZO[0-22]/LSMO[1-10], where mismatches are around -10% and 58%, respectively.

Figure 3 shows a cross-section STEM image projected on the $[001]_{op}/[110]_{ip}$ plane of LSMO. Two grains corresponding to [-211] and [2-1-1] HZO crystal variants can be recognized, while a sharp contrast between HZO and LSMO layers is visible. The contrast between atomic columns across the interface suggests, as seen in the HAADF image (which scales approximately as $Z^2$, Z being the atomic number),[39] that the semicoherent interface develops after 1 monolayer of pseudomorphic HZO. A similar monolayer was observed at the HZO/LSMO interface by Wei et al.[12] Interfacial layers, present in other oxide interfaces with large mismatch in lattice parameters and symmetry, have been proposed to be critical to allow the epitaxial growth.[40] Despite the huge mismatch f = -10%, the interface is coherent (see zooms in Figures 3 b-c). The zoomed view (Figure 3c) reveals the matching of n = 10 atomic La/Sr planes of the LSMO electrode with m = 9 atomic Hf/Zr planes of the HZO film. Note that the matching of n with m planes represents n and m times the distance between atomic columns; therefore, for n = 10 and m = 9, n+1=11 and m+1=10 columns are shown, respectively. Fast Fourier Transform (FFT) filtering of the images easily visualizes the additional planes, i.e. dislocations. In this case of grains with [-211] orientation, the domain size is deduced from the distance between these additional planes of LSMO, since they are more clearly visualized in the FFT filtered image than the location where best matching of HZO and LSMO planes occurs. The STEM image (Figure 3b) after FFT filtering (the FFT is presented in the inset) is shown in Figure 3d, displaying distances between adjacent additional planes of 9/10 and 10/11, indicating the presence of domains with same periodicities. Domains with m/n ratios 9/10 are found in all [-211] crystal variants examined, with presence of domains of close size, such as 6/7 or 11/12 (see S3, Supporting Information). About 3 or 4 of such domains fit within one single crystallite, given the domains and grain sizes, the latter being about 10-12 nm (see S3, Supporting Information).

The STEM characterization of HZO[0-22]/LSMO[1-10] interfaces is summarized in Figure 4. The cross-sectional STEM view in Figure 4a shows a [0-22] grain variant occupying most of the imaged HZO film. There is also a second [0-22] grain at its right. The zoom of the main [0-22] grain shows a semicoherent discontinuity. Here, the contrast between atomic columns across the interface also suggests, as in the [-211] grain presented in Figure 3, that the semicoherent interface develops after 1 monolayer of pseudomorphic HZO. The analysis of the semicoherent interface (Figure 4b-c) shows much smaller domains compared to [-211] variants. Here, most of the domains correspond to the matching of 3 atomic Hf/Zr planes and 2 La/Sr planes. The FFT filtered image (Figure 4c) also reveals the presence of an important number of 2/1 domains, appearing approximately 4 times less frequently than 3/2 domains.

The observed mechanism of DME of HZO(111) on LSMO(001) results in an effective lattice mismatch $f^*$ (%) = 100 x $(n \cdot d_{LSMO} - m \cdot d_{HZO})/n \cdot d_{HZO}$ that is much smaller than the lattice mismatch $f$ of direct accommodation of one-to-one lattice planes. In the case of HZO(111)/LSMO(001), the m/n values that minimize $f^*$ are 3/2 and 9/10 for HZO[0-22]/LSMO[110] and HZO[-211]/LSMO[110] interfaces, respectively (see Supporting Information S4). The residual mismatches after forming domains with these m/n are $f^*$ (HZO[0-22]/LSMO[110]) = 5.3 % and $f^*$ (HZO[-211] / LSMO[110]) = 0.03 %. The residual mismatch is still significant in the first case. However, domains with close m and n values can coexist to allow exact coincidence along a longer interface distance. In the case of the HZO[0-22]/LSMO[110] interface, a minimal residual mismatch can be achieved from the combination of about 25% of 2/1 domains with 75% of 3/2 domains. In the case of the HZO[-211]/LSMO[110] interface, most of the domains are 9/10, and the observed presence of domains producing tensile (6/7) or compressive (10/11 or 11/12) strain can combine to minimize the residual stress.

The formation of a semicoherent interface in HZO above a monolayer differs from the nucleation and glide of misfit dislocations above a critical film thickness that occurs in conventional epitaxy. Films grown by conventional epitaxy mechanism are elastically strained before plastic relaxation occurs, and the elastic energy, proportional to the thickness and the square of strain, can be high.[41] The stoichiometry of strained films can change locally[42] and other defects can form also to relieve the elastic energy,[43] and/or threading dislocations will be generated when the films relax plastically above the critical thickness.[44] These strain relief mechanisms can result in severe degradation of properties. In contrast, in films grown by DME mechanism, dislocations form at the interface when the film nucleates, permitting unstrained growth and thus without formation of strain-driven defects. When the film is cooled after growth, differences in thermal expansion coefficients between film and substrate, or a possible phase transition, would also introduce a small amount of residual elastic strain or some dislocations, but with much lower amount of strain-driven defects than in conventional high-mismatch epitaxy. Therefore, the advantage of DME is twofold: it permits epitaxy in largely mismatched systems and it results in a low-defect film. The occurring changes in surface symmetry, three-fold in HZO(111) and four-fold in LSMO(001), are similar to others that are frequent in heteroepitaxy.[34,45] The (111) growth of HZO occurs because of the lower HZO/LSMO interfacial energy compared to that between LSMO(001) and other low-index planes of HZO. Moreover, *ab-initio* calculations of the energy of HZO films of different orientations concluded that the (111)-orientation made the orthorhombic phase thermodynamically stable with respect to the monoclinic one.[46] Thus, the (111)-orientation would be optimal considering both HZO(111) film and HZO(111)/LSMO(001) interface energies. We finally remark that several contributions, interface energy, surface energy of o-HZO(111), and strain relief at domain and grain boundaries, are relevant to minimize the total energy in the complex epitaxy of o-HZO(111) on LSMO(001).

## Conclusions

In conclusion, it is demonstrated that epitaxy of orthorhombic HZO on LSMO(001) electrodes occurs by domain matching epitaxy. Scanning transmission electron microscopy shows domains of sizes that accommodate the lattice mismatch, and the formed orthorhombic HZO phase is of high crystal quality. The occurrence of domain matching epitaxy mechanism to stabilize the ferroelectric phase of $HfO_2$ opens the possibility to design epitaxial heterostructures oriented along other crystal directions or combining ferroelectric $HfO_2$ with other functional oxides.

## Experimental

Thin films deposition: epitaxial heterostructures integrated by HZO (top layer, t = 10 nm) and LSMO (bottom layer, t = 25 nm) were grown on $GdScO_3$(001). The heterostructures were deposited in a single process by pulsed laser deposition (KrF excimer laser). Detailed information on growth conditions and ferroelectric properties is reported elsewhere.[19]

<u>Structural characterization</u>: the crystal phases of HZO and epitaxial relationships were determined by X-ray diffraction using Cu Kα radiation. A Siemens D5000 diffractometer with point detector was used to measure symmetric 2θ scans. A Bruker D8, equipped with 2D detector Vantec 500, was used to acquire pole figures around o-HZO (-111) asymmetric reflections. Characterization of the interface was done by scanning transmission electron microscopy using a Nion UltraSTEM 200, operated at 200 kV and equipped with a 5th order Nion aberration corrector. High-angle annular dark field (HAADF) images of cross-sectional specimens were recorded as-viewed along the [110] zone axes of the $GdScO_3$ substrate.

<u>Ferroelectric characterization</u>: Top Pt contacts, 19 μm diameter and 20 nm thick, were deposited through a stencil mask by magnetron sputtering. Ferroelectric loops were measured using AixACCT TFAnalyser2000 platform at 1 kHz, using the dielectric leakage current compensation (DLCC) method to reduce the leakage current.[47,48]

**Supporting information**

Laue fringes around (111) HZO reflection in $\theta$-$2\theta$ scan (S1), top view of HZO(111) on LSMO(001) (S2), matching of planes in [-211] grains (S3), and residual mismatch for different domains (S4).

**Acknowledgements**

Financial support from the Spanish Ministry of Economy, Competitiveness and Universities, through the "Severo Ochoa" Programme for Centres of Excellence in R&D (SEV-2015-0496) and the MAT2015-73839-JIN (MINECO/FEDER, EU) and MAT2017-85232-R (AEI/FEDER, EU) projects, and from Generalitat de Catalunya (2017 SGR 1377) is acknowledged. IF acknowledges Ramón y Cajal contract RYC-2017-22531. SE acknowledges the Spanish Ministry of Economy, Competitiveness and Universities for his PhD contract (SEV-2015-0496-16-3) and its cofunding by the ESF. SE work has been done as a part of his Ph.D. program in Materials Science at Universitat Autònoma de Barcelona. The electron microscopy performed at ORNL was supported by the Materials Sciences and Engineering Division of Basic Energy Sciences of the Office of Science of the U.S. Department of Energy. We also thank Jaume Gázquez for his assistance with the STEM characterization.

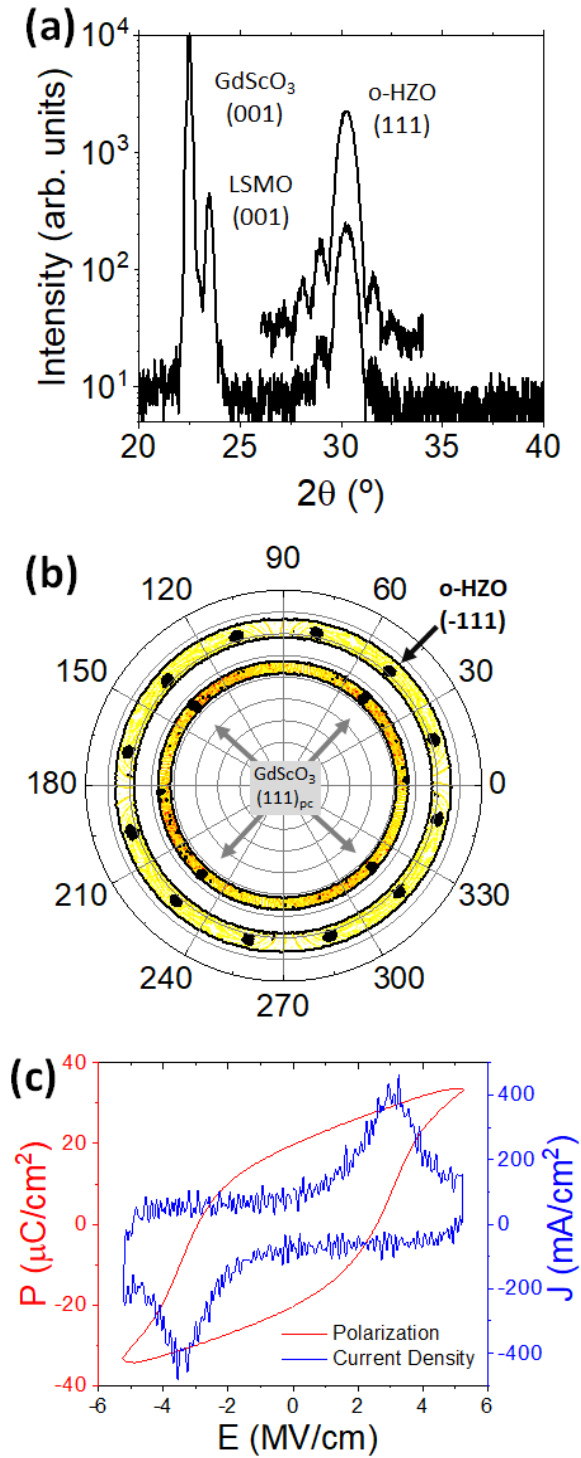

**Figure 1**. (a) XRD θ-2θ scan around symmetric reflections. The scan vertically shifted was recorded with longer time per step. (b) XRD pole figure around asymmetric GdScO₃(111) and o-HZO(-111) reflections. (c) Polarization loop at 1 kHz.

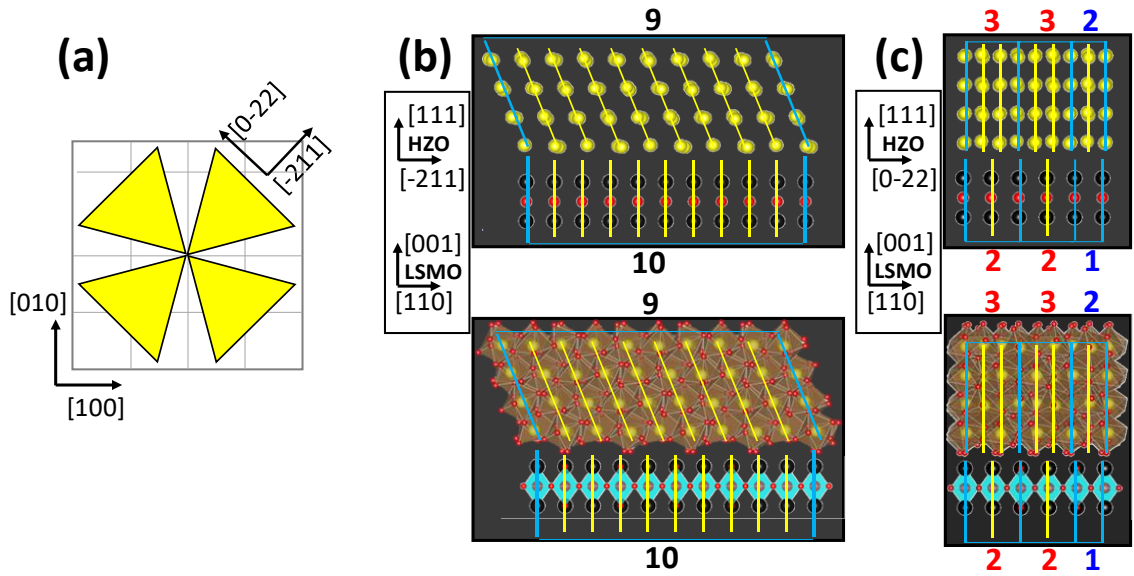

**Figure 2**. (a) Schematic showing a top view of HZO crystallites on the LSMO(001) surface. The 4 existing in-plane orientation variants of HZO crystallites are schematized with 4 yellow triangles. (b) and (c) Predicted optimal domains for [-211] and [0-22] HZO variants, respectively. For it, (-111) (for [-211] variants) and (0-22) (for [0-22] variants) HZO planes and LSMO(110) planes are marked. Only cations are shown in the sketches at the top, while both cations and oxygen atoms with the corresponding polyhedral sides are shown at the bottom. Planes delimiting the border of a domain are in light blue while planes inside a domain are in yellow. In (b), 9 (-111) matching 10 (110) planes is shown as the domain with smaller residual strain, while combinations of 3/2 (red) and 2/1 (blue) (0-22)/(110) matching of planes is shown in (c) for [0-22] variants.

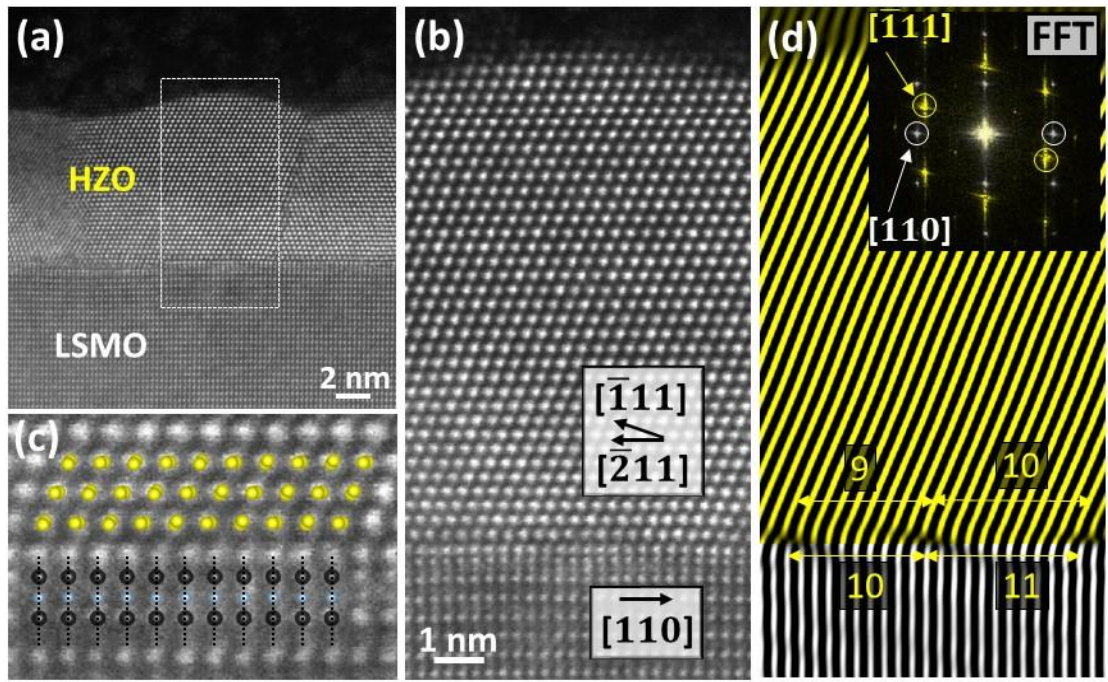

**Figure 3**. (a) Cross-sectional HAADF image of the HZO/LSMO heterostructure showing crystal variants of the [-211] type. (b) Zoom extracted from the region marked in (a). (c) Zoomed region around the interface with superimposed structural models of [-211] HZO and [110] LSMO, showing a 9/10 domain. (d) Reconstructed image from reflections in the Fourier space corresponding to [-111] HZO and [110] LSMO planes. The inset shows the FFT of both HZO and LSMO. For the sake of clarity, planes in the HZO layer are shown in yellow while planes in the LSMO are white. The same color code applies to the reflections present in the FFT inset. Two adjacent 9/10 and 10/11 domains are indicated.

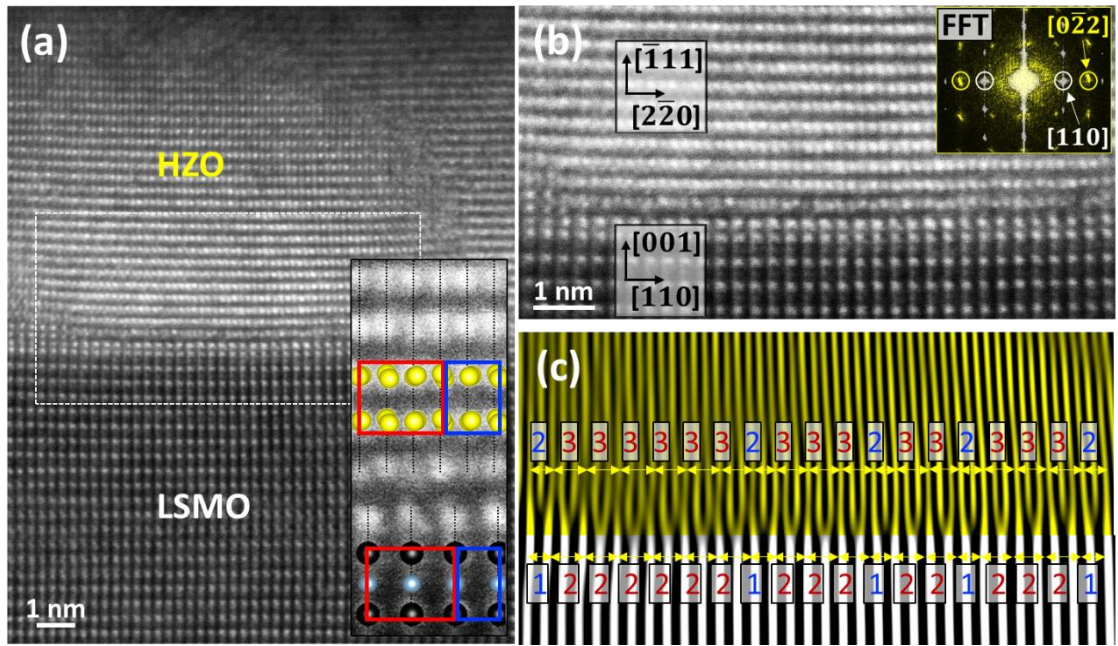

**Figure 4**. (a) Cross-sectional HAADF image of the HZO/LSMO heterostructure showing crystal variants of the [0-22] type. Bottom right inset shows a zoomed region around the interface with superimposed structural models of [0-22] HZO and [110] LSMO, showing two adjacent 3/2 (red) and 2/1 (blue) domains. (b) Zoom extracted from the region marked in white in (a) and FFT of HZO layer (yellow) and LSMO (white) in the top right inset. (c) Equivalent filtered image extracted from (b) by only considering the FFT marked reflections in the inset in (b). 3/2 (red) and 2/1 (blue) domains are visible. From first 2/1 domain on the left to last 3/2 on the right there is a total ratio of 18 domains (14 of which are 3/2 and 4 are 2/1), indicating that around a 77% of the present domains are 3/2 while around 23% are 2/1.

# Supporting information

# Domain Matching Epitaxy of Ferroelectric $Hf_{0.5}Zr_{0.5}O_2$(111) on $La_{2/3}Sr_{1/3}MnO_3$(001)

Saul Estandía,[1] Nico Dix,[1] Matthew F. Chisholm,[2] Ignasi Fina,[1] Florencio Sánchez[1,] *

[1] Institut de Ciència de Materials de Barcelona (ICMAB-CSIC), Campus UAB, Bellaterra 08193, Barcelona, Spain

[2] Center for Nanophase Materials Sciences, Oak Ridge National Laboratory, Tennessee 37831-6064, USA

* E-mail: fsanchez@icmab.es

## S1. Laue fringes around (111) HZO reflection in $\theta$-$2\theta$ scan

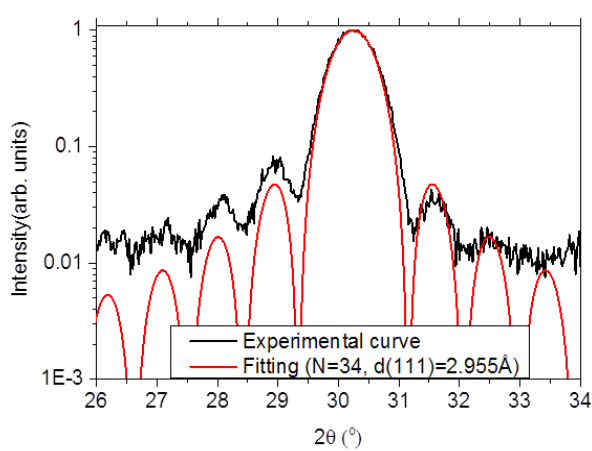

Figure S1. Experimental signal (black) and fitted model (red) of Laue fringes around the (111) reflection of HZO in $\theta$-$2\theta$ scan geometry. The following model equation was used

to fit the experimental data: $I = \frac{\sin^2(N \cdot d \cdot \frac{2\pi}{\lambda} \cdot \sin\theta)}{\sin^2(d \cdot \frac{2\pi}{\lambda} \cdot \sin\theta)}$ , where $\theta$ is the angle, $\lambda$ is the X-Rays wavelength, d is the out-of-plane plane spacing and N is the number of planes with d spacing. From its fitting, N=34 and d(111)=2.955 Å are obtained as best fit, indicating a total HZO thickness of N·d(111)=10.05 nm.

## S2. Top view of HZO(111) on LSMO(001)

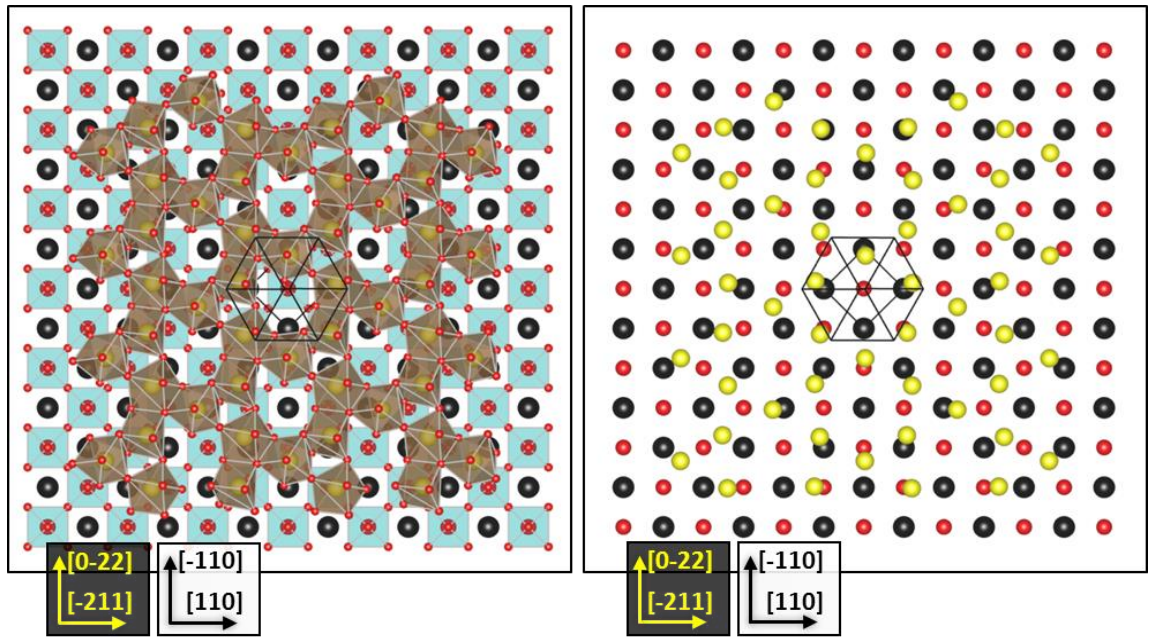

Figure S2. Left panel: Top view of first monolayer of (111)-HZO oxygen (red balls) forming polyhedra and cations (yellow balls) on the (001)-LSMO surface (La/Sr are black, while manganese and oxygen are red). Right panel: same top view as in left panel but showing only the cations. The visualization of the showed structures was done in VESTA.[1]

## S3. Matching of planes in [-211] grains

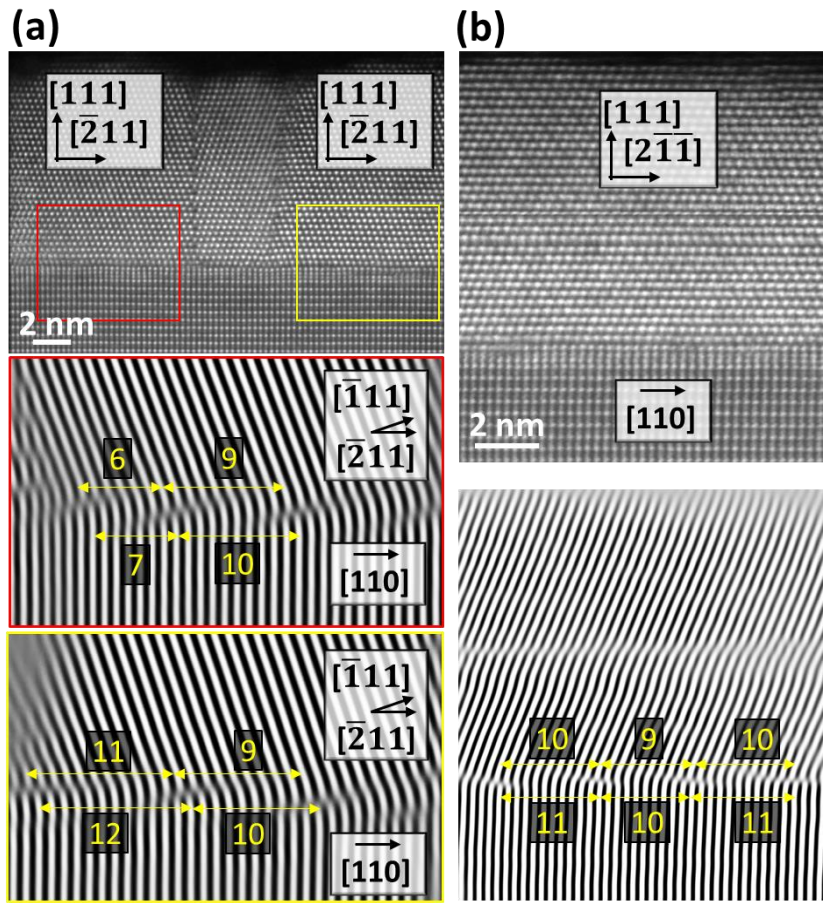

Figure S3. (a) In two other [-211] inspected grains, which are shown in the top panel, distances between additional planes indicate domains with m/n ratios above and below 9/10. Areas belonging to the two grains, marked in red and yellow, are analyzed in medium and bottom panels, respectively. (b) A bigger grain showing 4 misfit dislocations, and thus showing that up to 4 domains fit within a single grain.

## S4. Residual mismatch for different domains

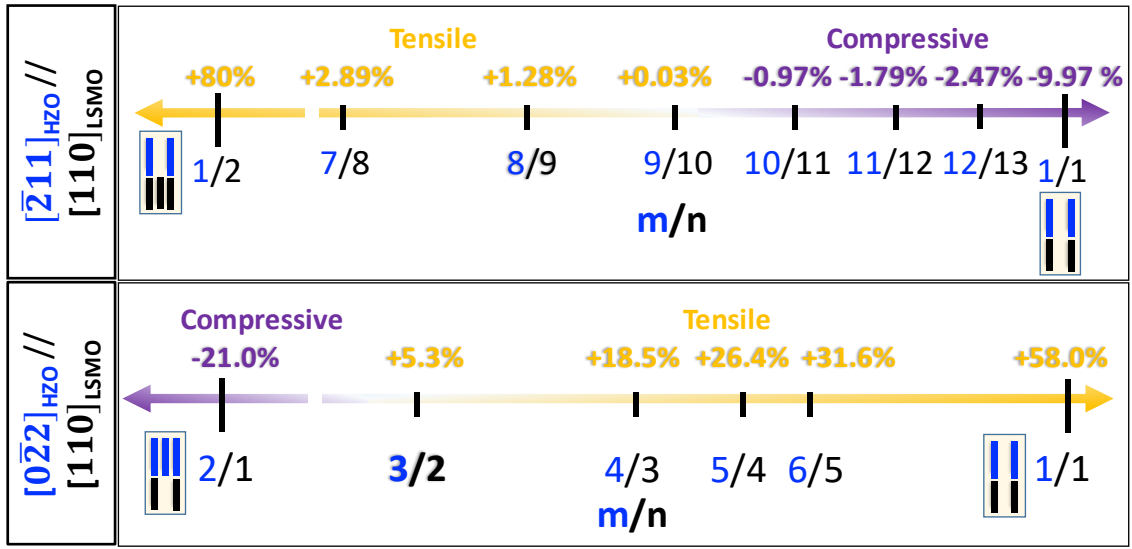

Figure S4. Residual mismatch for different m/n domains along [-211] direction (upper panel) and [0-22] (lower panel). Blue color is used to denote HZO planes, while black is employed for LSMO. The mismatch is calculated considered fully strained LSMO on GSO, that is, by using GSO lattice parameters. HZO cell parameters were taken from ref. 2.